\documentclass[prl,aps,twocolumn,10pt,superscriptaddress,notitlepage,longbibliography]{revtex4-2}

\usepackage[dvipsnames]{xcolor}
\usepackage[english]{babel} 
\usepackage{graphicx}
\usepackage{braket}
\usepackage{epstopdf}
\usepackage{mathtools}
\usepackage{amsmath}
\usepackage{amssymb}
\usepackage{bbm,bm}
\usepackage[caption=false]{subfig}
\usepackage{pythonhighlight}
\usepackage{hyperref}
\usepackage{float}
\usepackage{bbold}
\usepackage[T1]{fontenc}

\usepackage{mathtools}
\usepackage{dsfont}
\usepackage{bm}

\usepackage{tikz}
\usepackage{braket}

\usetikzlibrary{
    arrows.meta,
    positioning,
    calc,
    decorations.pathmorphing,
    shapes.geometric,
    fit
}

\graphicspath{./Figures/}

\usepackage[markup=underlined]{changes}
\makeatletter
\@namedef{Changes@AuthorColor}{magenta}
\colorlet{Changes@Color}{magenta}
\makeatother
\usepackage{hyperref}
 \hypersetup{
     colorlinks=true,
     linkcolor=blue,
     filecolor=blue,
     citecolor = magenta,      
     urlcolor=red,
     }

\usepackage{braket}
\usepackage{xargs}
\newcommandx{\greencom}[2][1=]
{\todo[inline, color=green!40,#1]{#2}}
\newcommandx{\bluecom}[2][1=]
{\todo[inline, color=blue!40,#1]{#2}}
\newcommandx{\bluemargin}[2][1=]
{\todo[color=blue!40,#1]{#2}}

\usepackage{letltxmacro}
\LetLtxMacro{\ORIGselectlanguage}{\selectlanguage}
\makeatletter
\DeclareRobustCommand{\selectlanguage}[1]{%
  \@ifundefined{alias@\string#1}
    {\ORIGselectlanguage{#1}}
    {\begingroup\edef\x{\endgroup
       \noexpand\ORIGselectlanguage{\@nameuse{alias@#1}}}\x}%
}
\newcommand{\definelanguagealias}[2]{%
  \@namedef{alias@#1}{#2}%
}

\makeatother

\definelanguagealias{en}{english}
\definelanguagealias{EN}{english}
\definelanguagealias{eng}{english}
\definelanguagealias{de}{ngerman}

\begin{document}
 
\title{A general no-go theorem for loss-free optical torque, and its enantiomer-differential escape route}
 
\author{Juanjuan Ren}
\affiliation{Department of Physics, Engineering Physics and Astronomy, Queen's University, Kingston, ON K7L 3N6, Canada}
\author{Reuven Gordon}
\affiliation{Department of Electrical and Computer Engineering,University of Victoria, Victoria, V8W 2Y2, Canada   }
\author{Stephen Hughes}
\affiliation{Department of Physics, Engineering Physics and Astronomy, Queen's University, Kingston, ON K7L 3N6, Canada}
 
\date{\today}
 
\begin{abstract}
We show that an isotropic, lossless dipolar particle, locally embedded in a homogeneous, lossless host, cannot experience axial optical torque under
any illumination, in any photonic environment, whatever its own chirality---a general identity
extending the sphere-in-free-space result to arbitrary chiral and substrate-coupled environments.
Shape anisotropy is the essential escape route, and the resulting torque is not generic:
closed-form, radiative-reaction-dressed polarizabilities for a chiral dielectric ellipsoid show
its loss-free ``optical spanner'' torque changing by an order of magnitude between enantiomers
($\kappa\to-\kappa$), a mechanical response impossible for any sphere. Above a high-index or lossy substrate like gold or silicon,
standing-wave enhancement, or simply a larger particle at a near-contact gap, brings it beyond
%standard 
optical-torque-wrench thresholds. 
\end{abstract}

\maketitle
 
Optical forces and torques, generated by exchanging linear and angular momentum between light and
matter, are used extensively to control the translational and rotational motion of micro- and
nanoparticles---trapping, levitation, transport, and
sorting~\cite{ashkin_acceleration_1970,ashkin_observation_1986,grier_revolution_2003,marago_optical_2013,gao_optical_2017,toftul_radiation_2026}.
Optical torque, in particular, transfers angular momentum to a particle: spin angular momentum
drives rotation about the particle's own axis, while orbital angular momentum and custom
wavefronts enable orbital trajectories~\cite{shen_optical_2019,he_review_2025}. 
The net torque is set jointly by the illumination and the particle's
response---absorption, anisotropy, and multipolar coupling to structured
fields~\cite{nieto-vesperinas_optical_2015,he_review_2025,toftul_radiation_2026}.

Particle chirality adds a magnetoelectric cross-coupling: opposite enantiomers (mirror-image
forms distinguished by the sign of $\kappa$) respond differently to identical illumination,
motivating enantioselective trapping, separation, pulling, and rotational
control~\cite{canaguier-durand_mechanical_2013,tkachenko_optofluidic_2014,fernandes_optical_2015,mun_electromagnetic_2020,kakkanattu_review_2021}.
A recently reported intensity-gradient torque on an \emph{absorptive} chiral sphere in a standing
wave is sign-set by chirality~\cite{wen_optical_2025}.
A planar interface adds further control, as surface feedback reshapes optical momentum and
angular-momentum exchange~\cite{quidant_surface-plasmon-based_2008,shi_advances_2023,toftul_radiation_2026}.
Wang and Chan showed a lossless chiral particle near a lossy film converts linear illumination
into a chirality-signed \emph{lateral} force~\cite{wang_lateral_2014}; an evanescent wave
introduces its own chirality-dependent lateral force~\cite{hayat_lateral_2015}, and bidirectional
lateral sorting of chiral Mie particles has been demonstrated
experimentally~\cite{shi_chirality-assisted_2020}. These works address the
interface-\emph{parallel} force; in contrast, the interface-\emph{normal} torque---the observable most tied to
particle spin---has received very little attention.
 
A more basic question underlies all of this: can a strictly \emph{lossless} particle experience
any axial torque at all? For an isotropic sphere in free space the answer has long been {\it no}---its
rotational symmetry keeps every multipole channel of its T-matrix separately unitary, so
extinction and recoil angular momentum cancel channel by
channel~\cite{marston1984,nieto-vesperinas_optical_2015,toftul_radiation_2026}.

In this Letter, we show that this null result is far more general than the free-space case in
which it was established: it holds \emph{exactly}, at dipole order, for an isotropic particle of
\emph{any} chirality coupled self-consistently to \emph{any} photonic environment---reciprocal or
not, lossy or lossless, achiral or bulk-chiral, planar or arbitrarily structured---because the
environment enters the proof only through the local field it produces at the particle, never
through its geometry or symmetry. { Shape anisotropy is therefore not one option among several
for generating loss-free torque; it is the {\it only} one available at dipole order, in any
environment.}

\begin{figure}[t]
\centering
\includegraphics[width=\columnwidth]{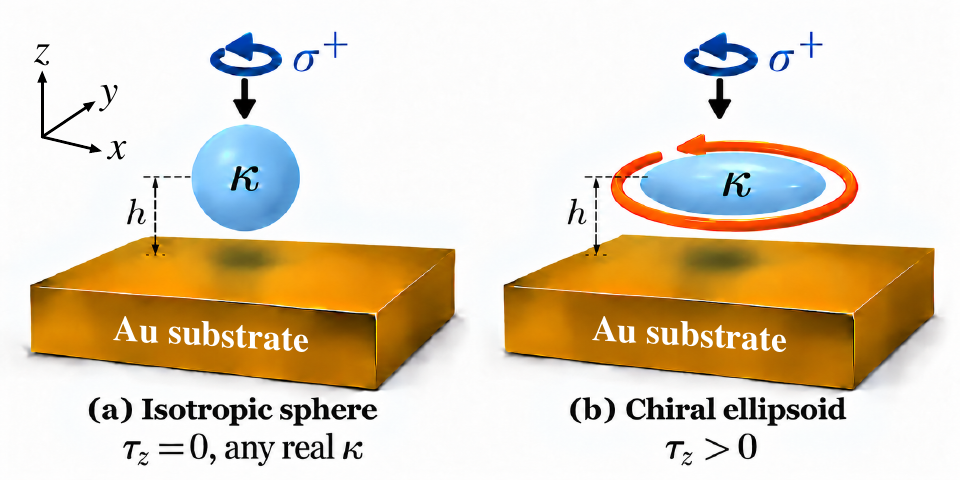}
\caption{
(a) A {\it lossless}, isotropic particle above a planar substrate has exactly zero axial torque
under any illumination, for any real chirality $\kappa$ [Eq.~\eqref{eq:tau}]. (b) Breaking isotropy
with an ellipsoidal shape gives the same {\it lossless} particle a genuine torque, $\tau_z>0$ for
either enantiomer but differing in magnitude between them [Eq.~\eqref{eq:leading}]. Both panels are
illuminated from above at normal incidence by circularly polarized light of handedness $\sigma^+$;
the ellipsoid's long axis lies in the substrate plane, perpendicular to the beam.}
\label{fig:schematic}
\end{figure}

We derive closed-form, radiative-reaction-dressed polarizabilities for a chiral dielectric
ellipsoid that make this explicit, use them to characterize how chirality modulates the resulting
shape-birefringence (``optical spanner'') torque near a planar gold interface
(Fig.~\ref{fig:schematic}), and show that a height-optimized, non-contact geometry brings a
realistic particle---and hence anisotropic, chirality-transducing---within reach of torque-wrench detection.
The same symmetry protection applies to an isotropic, achiral emitter's decay rate, which is
likewise shielded from enantiomer-differential Purcell shifts until shape- or environment-side
anisotropy unlocks it~\cite{guzatov2018,butcher2012,rapp2025}.

\textit{No-go theorem.}---Consider a small, isotropic (spherical), lossless particle characterized
by a bare $6\times6$ bi-isotropic (i.e., isotropic magnetoelectric) polarizability
(given in the End Matter), coupled self-consistently to \emph{any} photonic
environment via an electromagnetic Green's function $\mathbf{G}$ through $\tilde{\boldsymbol{\alpha}}^{-1}=\boldsymbol{\alpha}^{-1}-\mathbf{G}$. Then the axial
optical torque vanishes exactly (with $z$ along the illumination axis in the absence of a
substrate),
since
\begin{equation}
\tau_z \equiv \tfrac12\mathrm{Re}\big[p_x^*E_y-p_y^*E_x\big]+\tfrac12\mathrm{Re}\big[m_x^*H_y-m_y^*H_x\big] = 0,
\label{eq:tau}
\end{equation}
for \emph{any} $\mathbf{G}$ (reciprocal or not, lossy or not, axially symmetric or not), any illumination
polarization, and any particle chirality $\kappa$ (real). The proof is short: for an isotropic lossless
particle, 
$\mathbf p=\alpha_e\epsilon_0\epsilon_B\mathbf E+i\alpha_\chi\sqrt{\epsilon_0\epsilon_B\mu_0\mu_B}\,\mathbf H$ and
$\mathbf m=-i\alpha_\chi\sqrt{\epsilon_0\epsilon_B\mu_0\mu_B}\,\mathbf E+\alpha_m\mu_0\mu_B\mathbf H$ ($\mathbf{m}$
is in $\rm Wb\cdot m$ throughout) with real bare  $\alpha_e,\alpha_m,\alpha_\chi$, so
$\mathbf p^*\times\mathbf E+\mathbf m^*\times\mathbf H
=\alpha_e\epsilon_0\epsilon_B\mathbf E^*\times\mathbf E+\alpha_m\mu_0\mu_B\mathbf H^*\times\mathbf H
+2i\alpha_\chi\,\sqrt{\epsilon_0\epsilon_B\mu_0\mu_B}\,\mathrm{Re}(\mathbf E^*\times\mathbf H)$ is purely imaginary for \emph{whatever}
total field $(\mathbf E,\mathbf H)$ the particle sees, thus Eq.~\eqref{eq:tau} vanishes identically
regardless of $\mathbf{G}$ or the illumination that generated that field
(the identity holds component-by-component, so the full torque vector vanishes, not only its
axial component; we focus on $\tau_z$ throughout since it is the observable measured by a
substrate-coupled torque wrench).
Physically, this is the statement
that {\it a lossless isotropic dipole scatterer cannot absorb angular momentum}: extinction and recoil torques cancel exactly. 

For a lossless isotropic sphere of arbitrary size, the same conclusion follows from the unitarity and rotational invariance of its intrinsic scattering matrix $S$ in the homogeneous, isotropic, lossless host,
provided a shell of this host of nonzero thickness surrounds the sphere. With $S$ defined on flux-normalized spherical-wave channels, $S^\dagger S=I$ and $[S,J_i]=0$ for $i=x,y,z$. 
Hence, for arbitrary incoming fields, including all  environmental feedback, the angular-momentum flux is balanced in all three Cartesian components about the sphere center, and the time-averaged torque vanishes.
This covers, in particular, a sphere in water/solution above a substrate, provided a nonzero solution gap
separates the two: the substrate may be chiral, lossy, anisotropic, or structured, and the
illumination arbitrary.

We stress that $\mathbf p$, $\mathbf m$, and the total field entering Eq.~\eqref{eq:tau} are
\emph{exact}: the self-consistently dressed moments and full local field obtained from
$\mathbf{u}=(\mathbf{I}-\boldsymbol{\alpha} \mathbf{G})^{-1}\boldsymbol{\alpha}\, \mathbf{d}$ 
with $\mathbf{u}=(\mathbf p/\sqrt{\epsilon_0\epsilon_B},\,\mathbf m/\sqrt{\mu_0\mu_B})$ and
$\mathbf{d}=(\sqrt{\epsilon_0\epsilon_B}\,\mathbf E_0,\,\sqrt{\mu_0\mu_B}\,\mathbf H_0)$, 
at arbitrarily strong coupling, not the bare polarizability acting on the excitation field alone.
This distinction is what makes the proof possible: the \emph{bare}
$\alpha_e,\alpha_m,\alpha_\chi$ are real for a lossless material, whereas the \emph{dressed}
$\tilde{\boldsymbol{\alpha}}=(\boldsymbol{\alpha}^{-1}-\mathbf{G})^{-1}$ is generically complex,
since $\mathbf{G}$ carries an imaginary (radiative) part; writing the argument in terms of
$\tilde{\boldsymbol{\alpha}}$ and $\mathbf E_{0}$ would obscure the cancellation entirely.

What is new relative to the classic free-space
result~\cite{marston1984,bohren1974} is the statement, and proof, for arbitrary environments,
which follows from the locality of the dipole torque: the identity depends only on the total field at the particle's own
position, not on any global property of the environment that produced it. A large \emph{differential} coupling
strength between circular-polarization channels does not, by itself, confer any capacity for torque
on an isotropic lossless particle, however large the differential: escaping the theorem requires particle absorption or breaking the
rotational invariance of the particle's intrinsic response, via shape anisotropy (this work; also via
discrete/multipolar symmetry~\cite{achouri2023}), or material anisotropy.

% Illumination geometry alone does not help, even off-axis. Wang and Chan~\cite{wang_lateral_2014}
% report a lateral force on a lossless chiral sphere under oblique/evanescent illumination near a
% surface, and note---without calculating it---that an accompanying enhanced torque should exist; the
% identity underlying Eq.~\eqref{eq:tau} in fact holds component-by-component,
% \blue{OK, this is mentioned here. Then I'm ok if we only focus on axial torque earlier.}
% so this torque must
% vanish exactly in the strictly lossless limit, for {\it any} illumination direction or substrate
% coupling, resolving their conjecture. The protection is nonetheless specific to torque, not to optical
% momentum transfer in general: a lateral (recoil) \emph{force} \emph{can} act on an isotropic,
% achiral particle near a surface from the illumination's spin alone, with no particle chirality
% required~\cite{rodriguezfortuno2015}.

Illumination geometry alone does not help, even off-axis: Wang and Chan~\cite{wang_lateral_2014}
conjecture an enhanced torque accompanying their lateral force on a lossless chiral sphere under
oblique/evanescent illumination, but Eq.~\eqref{eq:tau} holds component-by-component, so that
torque must vanish exactly in the strictly lossless limit, for {\it any} illumination direction or substrate coupling, resolving their conjecture. The protection is specific to torque: a
lateral (recoil) \emph{force} \emph{can} act on an isotropic, achiral particle near a surface from
the illumination's spin alone~\cite{rodriguezfortuno2015}.

\textit{Shape anisotropy: the escape route.}---For a dielectric, optically chiral (Pasteur)
ellipsoidal particle with permittivity $\epsilon_p$, permeability $\mu_p$, chirality parameter $\kappa$, and volume $V_p$
in a background of permittivity $\epsilon_B$, permeability $\mu_B$,  the bare polarizability along each transverse
principal axis $i=1,2$ (depolarization factor $L_i$) follows from
$D_{e,i}\equiv\epsilon_p-\epsilon_B+\epsilon_B/L_i$ and $D_{m,i}\equiv\mu_p-\mu_B+\mu_B/L_i$ as
$\alpha_{e,i}=(V_p/L_i)[D_{m,i}(\epsilon_p-\epsilon_B)-\kappa^2]/\mathcal{D}_i$,
$\alpha_{m,i}=(V_p/L_i)[D_{e,i}(\mu_p-\mu_B)-\kappa^2]/\mathcal{D}_i$, and
$\alpha_{\chi,i}=\kappa V_p\sqrt{\epsilon_B\mu_B}/(L_i^2\,\mathcal{D}_i)$, with
$\mathcal{D}_i\equiv D_{e,i}D_{m,i}-\kappa^2$ (a time-harmonic factor $e^{-i\omega t}$ is used
throughout).

Dressing with the environment's Green's function $\mathbf{G}$ through $\tilde{\boldsymbol{\alpha}}^{-1}=\boldsymbol{\alpha}^{-1}-\mathbf{G}$
[the same relation entering Eq.~\eqref{eq:tau}] couples the electric, magnetic, and
magnetoelectric channels self-consistently. In a homogeneous medium, $\mathbf{G}=ig_B\mathbf{I}_4$ with
$g_B=k_B^3/6\pi$ the homogeneous medium radiative-reaction rate ($k_B=\sqrt{\epsilon_{B}\mu_B}\omega/c$), and the $4\times4$ problem decouples
exactly into the two transverse axes; with $\Delta_i\equiv\alpha_{e,i}\alpha_{m,i}-\alpha_{\chi,i}^2$,
the exact self-consistent (radiative-reaction-dressed) polarizabilities are
\begin{align}
\tilde\alpha_{e,i} &= \frac{\alpha_{e,i}-ig_B\Delta_i}{1-ig_B(\alpha_{e,i}+\alpha_{m,i})-g_B^2\Delta_i}, \nonumber\\
\tilde\alpha_{m,i} &= \frac{\alpha_{m,i}-ig_B\Delta_i}{1-ig_B(\alpha_{e,i}+\alpha_{m,i})-g_B^2\Delta_i}, \nonumber\\
\tilde\alpha_{\chi,i} &= \frac{\alpha_{\chi,i}}{1-ig_B(\alpha_{e,i}+\alpha_{m,i})-g_B^2\Delta_i},
\label{eq:dressed}
\end{align}
which reproduce the exact coupled-dipole solution to machine precision. To our knowledge such
closed-form dressed polarizabilities have not been given before for a shape-anisotropic chiral
particle, the only comparable treatment being the spherical case~\cite{golat_optical_2024}.

With $\Delta_e\equiv\alpha_{e,1}-\alpha_{e,2}$ (similarly $\Delta_m,\Delta_\chi$), the leading-order
torque under normally-incident, circularly polarized illumination of amplitude $E_{inc}$~\footnote{All results use right circularly polarized (RCP, $\sigma^+$) illumination,
$\mathbf {E}_{inc}={E_{inc}}(\hat x+i\hat y)$ with $\mathbf k\parallel-\hat z$, corresponding to clockwise
$\mathbf E$-field rotation as seen looking into the oncoming beam. 
Under left circularly polarized (LCP, $\sigma^-$) illumination, $\tau_z(\kappa,\sigma^-)=-\tau_z(-\kappa,\sigma^+)$
exactly, so LCP results follow directly from the $\kappa\to-\kappa$ dependence already shown
throughout.},  is
\begin{equation}
\tau_z\approx\tfrac12\epsilon_0\epsilon_Bg_BE_{inc}^2\big[(\Delta_e-\Delta_\chi)^2+(\Delta_m-\Delta_\chi)^2\big],
\label{eq:leading}
\end{equation}
manifestly non-negative and vanishing identically in the isotropic (sphere) limit
$\Delta_e=\Delta_m=\Delta_\chi=0$---the fingerprint of the no-go theorem above. This is the
familiar shape-birefringence ``optical spanner'' effect~\cite{friese1998}; the new ingredient is its
{\it closed-form chirality dependence}, Eqs.~\eqref{eq:dressed} and \eqref{eq:leading}, and the substrate extension below.
Being $O(g_B)$, it is the leading-order sum of extinction and recoil contributions, which can
partially cancel; it is not an absorption torque.% (End Matter).

The enantiomer asymmetry is already visible here. From the bare polarizabilities above, $\Delta_e$ and
$\Delta_m$ are even in $\kappa$ while $\Delta_\chi$ is odd, so $\kappa\to-\kappa$ sends
$(\Delta_e-\Delta_\chi)^2\to(\Delta_e+\Delta_\chi)^2$ and, to leading order in $g_B$,
$\tau_z(\kappa)-\tau_z(-\kappa)\approx-2\epsilon_0\epsilon_Bg_BE_{inc}^2\Delta_\chi(\Delta_e+\Delta_m)$---a
\emph{cross} term, first order in $\kappa$ rather than $\kappa^2$ and nonzero only when
$\Delta_\chi(\Delta_e+\Delta_m)\neq0$. Chirality and transverse shape anisotropy are therefore both
required, and the response is a magnitude change rather than a sign flip.

\textit{Coupling to a photonic environment.}---In the ordered basis $(p_x,p_y,m_x,m_y)$, the bare
polarizability is the $4\times4$ matrix $\boldsymbol{\alpha}=\mathrm{diag}(\alpha_{e,1},\alpha_{e,2},\alpha_{m,1},\alpha_{m,2})$
plus off-diagonal magnetoelectric blocks $\pm i\,\mathrm{diag}(\alpha_{\chi,1},\alpha_{\chi,2})$
(End Matter),
and the self-consistent moments
follow from $\mathbf{u}=(\mathbf{I}_4-\boldsymbol{\alpha}\mathbf{G})^{-1}\boldsymbol{\alpha}\mathbf{d}$, with $\mathbf{d}$ the (possibly substrate-reflected) excitation
field and $\mathbf{G}$ the environment's coupling matrix;
the homogeneous medium case, $\mathbf{G}=ig_B\mathbf{I}_4$, is what
produces Eq.~\eqref{eq:dressed}. Above a planar, isotropic, reciprocal, achiral, nonmagnetic substrate at height $h$, normal
incidence keeps $p_z=m_z=0$ 
%exactly 
(an exact decoupling of the $6\times6$ problem, not a
truncation, when two of the ellipsoid's principal axes are in-plane), and $\mathbf{G}$ acquires near-field terms $G_{ee}(h)$, $G_{mm}(h)$ and a magnetoelectric term $G_{em}(h)$. 

 With
$a\equiv ig_B{+}G_{ee}(h)$, $b\equiv G_{em}(h)$, and $c\equiv ig_B{+}G_{mm}(h)$, 
\begin{equation}
\mathbf{G}=\begin{pmatrix} a\mathbf{I}_2 & b\mathbf{J}\\ b\mathbf{J} & c\mathbf{I}_2 \end{pmatrix},
\qquad
\mathbf{J}=\begin{pmatrix} 0 & -1\\ 1 & 0 \end{pmatrix},
\label{eq:Gmatrix}
\end{equation}
where this $\mathbf{G}$ is reciprocal---the off-diagonal blocks are transposes of each other up to a
sign, which with $\mathbf J^{\rm T}=-\mathbf J$ is why both carry the same scalar $b$---and is
verified independently against perfect-conductor image dipoles and direct angular-spectrum
integration. Also note that $G_{mm}(h)$ is much smaller than $G_{ee}(h)$ as $k_Bh\to0$, but is \emph{not} small
relative to $g_B$ itself, and is retained throughout (see End Matter).

% writing
% $a\equiv ig_B{+}G_{ee}(h)$, $b\equiv G_{em}(h)$, $c\equiv ig_B{+}G_{mm}(h)$, and
% $\mathbf{J}\equiv\left(\begin{smallmatrix}0&-1\\1&0\end{smallmatrix}\right)$,
% \begin{equation}
% \mathbf{G}=\begin{pmatrix} a\mathbf{I}_2 & b\mathbf{J}\\ b\mathbf{J} & c\mathbf{I}_2 \end{pmatrix},
% \label{eq:Gmatrix}
% \end{equation}
% reciprocal 
% (two off-diagonal blocks are
% transposes of each other up to a sign, which with
% $\mathbf J^{\rm T}=-\mathbf J$ is why both carry the same scalar $b$), verified independently against perfect-conductor image-dipoles and direct angular-spectrum integration; $G_{mm}(h)$ is much smaller than $G_{ee}(h)$ as
% $k_Bh\to0$, but is \emph{not} small relative to $g_B$ itself, and is retained throughout (see End
% Matter).

In the near-field ($k_Bh\ll1$) limit, $G_{ee}$ reduces to the standard image-dipole form~\cite{chanceprocksilbey1978,barnes1998}, whose validity domain in a coupled-dipole
setting is characterized in Ref.~\cite{gay-balmaz_electromagnetic_2001}, and $G_{em}$ to its leading, $O(k_Bh)$, magnetoelectric correction,
\begin{equation}
G_{ee}(h)=-\frac{\eta(\omega)}{32\pi h^3}, \qquad
G_{em}(h)=+i\,\frac{\eta(\omega)k_B}{32\pi h^2}, 
\label{eq:Gnearfield}
\end{equation}
where $\eta\equiv-(\epsilon_S-\epsilon_B)/(\epsilon_S+\epsilon_B)$
with $\epsilon_S$  the (complex, dispersive) substrate permittivity. For a lossless particle under circular polarization above a planar substrate, 
however, the torque is carried entirely by the \emph{anti-Hermitian} (radiative plus absorptive)
part of this coupling [$\mathrm{Im}\,[G_{ee}]$, $\mathrm{Im}\,[G_{mm}]$ and, the block being antisymmetric,
$\mathrm{Re}\,[G_{em}]$; the first dominates], which Eq.~\eqref{eq:Gnearfield} cannot capture quantitatively once
$k_Bh$ is not vanishingly small: we therefore solve self-consistently with the full retarded
(Sommerfeld) reflected Green tensor throughout this
work~\cite{sommerfeld1909,paulus2000,novotnyhecht2025}, obtained by direct
numerical angular-spectrum integration over the in-plane wavevector $k_\rho$ (explicit integral
forms, numerical validation, and gold's optical constants~\cite{johnson1972} are given in the End
Matter).

\begin{figure}[!t]
\centering
\includegraphics[width=\columnwidth]{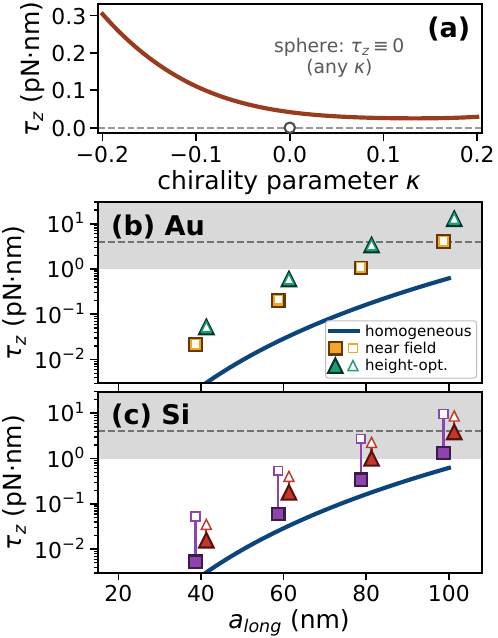}
\caption{
(a) Homogeneous-medium torque versus chirality parameter $\kappa$ for the $60$/$24$~nm semi-axis
ellipsoid; the sphere's exact zero is marked. (b),(c) Torque versus long semi-axis $a_{long}$
(aspect ratio $2.5$) in a homogeneous medium at $\kappa={+}0.2$ (blue curve), and from the exact
Sommerfeld substrate solve for both enantiomers ($\kappa={+}0.2$ filled, ${-}0.2$ open) at a
near-contact gap ($d_{gap}=2$~nm) and at the height-optimized antinode, above (b) gold and (c)
silicon. Shading marks a representative detectability floor, $\tau\gtrsim1$~pN$\cdot$nm, with
$k_{Bolt}T$ dashed. 
On gold the two enantiomers differ by only a few percent, so the filled and open
markers overlap; on silicon their torques differ by a factor of 7--10.
}
\label{fig:results}
\end{figure}

\textit{Chirality modulation.}---Figure~\ref{fig:results}(a)
shows the exact homogeneous-medium torque [Eqs.~\eqref{eq:dressed}--\eqref{eq:leading} solved
self-consistently, without approximation] versus $\kappa$ for a water-immersed ellipsoid of
semi-axes $a_{long}=60$~nm, $a_{short}=24$~nm (a $120\times48$~nm particle), with
$\epsilon_p=1.45^2$, $\lambda_0=1\,\mu$m, and per-component amplitude $E_{inc}=3\times10^8$~V/m. The response is strongly asymmetric in $\kappa$: reversing the sign from
$\kappa=+0.2$ to $\kappa=-0.2$ changes $\tau_z$ by an order of magnitude (from $0.029$ to
$0.304$~pN$\cdot$nm) rather than simply flipping sign, exactly as the cross term linear in $\Delta_\chi$ requires---an
{\it enantiomer-differential signature} that is {\it structurally forbidden for any lossless sphere by
Eq.~\eqref{eq:tau}}, and mechanically analogous to the shape-enhanced decay-rate discrimination of
Ref.~\cite{guzatov2018}.

\textit{Substrate enhancement and detectability.}---Because the local excitation field is itself a
standing wave formed with the reflected beam, the torque need not peak at the smallest achievable
gap: for a $60$~nm ellipsoid above gold with an enhanced (metamaterial-scale) chirality
$\kappa=0.2$, the exact solve gives $0.20$~pN$\cdot$nm at a $2$~nm gap but $0.60$~pN$\cdot$nm at
$h\approx185\,$nm $\approx\lambda_0/4n_B$, a standing-wave antinode nearly $3\times$ larger.
Repeating the height-optimized solve at $a_{long}=40$, $80$, and $100$~nm reproduces this
behavior and lets the torque grow steeply with size: Fig.~\ref{fig:results}(b) places these
exact values against the
homogeneous-medium size scaling and against a representative detectability floor,
$\tau\gtrsim1$~pN$\cdot$nm, typical of birefringent-particle optical-torque wrenches, with the
thermal scale $k_{Bolt}T\approx4$~pN$\cdot$nm (with $k_{Bolt}$ the Boltzmann constant) at room temperature marked for
reference~\cite{gutierrezmedina2010,pedaci2012}.

Notably, the $80$ and $100$~nm particles clear this floor at the near-contact gap alone ($1.06$ and
$4.00$~pN$\cdot$nm respectively), without the standing-wave enhancement at all, at an intensity
$I=c\epsilon_0n_BE_{inc}^2=3.2\times10^{10}$~W/cm$^2$. The torque scales linearly with intensity, so
standard continuous-wave torque-wrench operating intensities ($\sim10^6$--$10^7$~W/cm$^2$) would
place these configurations below detection at equal particle size---a conservative reference,
because the torque is constant (DC): orientation accumulates linearly in the integration time
while thermal rotational diffusion grows only as its square root, so the signal-to-noise ratio
improves with averaging, as in single-molecule torque measurements~\cite{pedaci2012}. A
circular-dichroism-style protocol---comparing accumulated rotation under RCP versus LCP
illumination---isolates the chirality-driven signal from achiral background and drift; 
closing the remaining gap otherwise likely requires tightly focused or pulsed illumination, or a slightly lossy particle.

% closing the
% remaining gap otherwise likely requires tightly focused or pulsed illumination, or a small
% departure from strict losslessness.

The size parameter $k_Ba_{long}$ ranges from $0.25$ at $30$~nm to $0.84$ at $100$~nm, where quadrupole corrections are not
strictly negligible; the weak index contrast throughout ($n_p/n_B\approx1.09$) 
keeps the electric-dipole response dominant over a wider size range than a strongly scattering or metallic particle would allow, but we regard
 the $80$ and $100$~nm values as leading-order
dipole estimates. The no-go theorem itself, Eq.~\eqref{eq:tau}, holds at any sphere size and is
unaffected.

\textit{Role of the substrate: gold versus silicon, and the surface plasmon.}---Gold's large,
predominantly real, negative permittivity ($\epsilon_{\rm Au}=-41.8+2.9i$ at
$\lambda_0=1\,\mu$m~\cite{johnson1972}) also supports a surface-plasmon pole in $r_p(k_\rho)$, so it
is natural to ask whether the enhancement above is a resonant plasmonic effect; our
angular-spectrum integration locates and resolves this pole explicitly (End Matter).
Repeating the
full calculation for a near-lossless, high-index dielectric substrate
[silicon, $\epsilon_{\rm Si}=12.76+0.004i$~\cite{green2008}, whose positive $\mathrm{Re}\,[\epsilon_S]$
admits no surface-plasmon pole] gives $0.060$~pN$\cdot$nm at $h=26$~nm and
$0.18$~pN$\cdot$nm at $h=185$~nm---smaller than gold by a factor of $3.3$--$3.4$ at both heights,
a remarkably uniform ratio across a $7\times$ change in height that is not the signature of a
narrowband resonant enhancement. 

Substrate dissipation is likewise not what drives the height-optimized enhancement:
setting gold's loss artificially to zero at fixed $\mathrm{Re}\,\epsilon_{\rm Au}$ changes
the torque by only $\sim10\%$ for $h\gtrsim140$~nm, although at a near-contact gap
($d_{gap}=2$~nm) the same test changes it by $38$--$56\%$ at $\kappa=0.2$, where
dissipative near-field channels do contribute (End Matter). Gold's
advantage over silicon is therefore set almost entirely by its much larger, \emph{broadband}
reflectivity---acting as a near-ideal mirror across the full relevant range of $k_\rho$,
rather than by resonant coupling into the narrowband plasmon
mode.

% A more direct test confirms this: retaining gold's large negative real permittivity but setting
% its loss artificially to zero changes the torque by only $\sim10\%$ at every height tested---far
% too little for a genuinely lossy resonant channel to be responsible, since a real surface plasmon
% is by definition a lossy mode (End Matter). Gold's
% advantage over silicon is therefore set almost entirely by its much larger, \emph{broadband}
% reflectivity---acting as a near-ideal mirror across the full relevant range of $k_\rho$, 
% %largely independent of whether it absorbs---
% rather than by resonant coupling into the narrowband plasmon
% mode.

This same broadband reflectivity, however, is also what nearly erases the  enantiomer-differential signal on gold: Fig.~\ref{fig:results}(b) shows $\kappa={\pm}0.2$ within
a few percent of each other at every size, versus the $10.5\times$ separation in the homogeneous
medium [Fig.~\ref{fig:results}(a)]. Silicon's much weaker reflectivity keeps the local excitation
closer to a traveling wave, and Fig.~\ref{fig:results}(c) shows the enantiomer separation
recovering to a $7$--$10\times$ contrast at $d_{gap}=2~$nm, comparable to the
homogeneous-medium result, while still clearing the detectability floor at $a_{long}=100$~nm. Gold therefore maximizes the achiral torque baseline, while silicon---despite
$\sim3.3\times$ smaller torque at $\kappa=0.2$---preserves the chirality-differential signature that is this work's central observable; which substrate is preferable depends on whether peak torque or enantiomer discrimination is the experimental goal.

A bulk-chiral substrate adds a further, independent modulation channel, 
first order in $\kappa_S$ and connected to the chiral Casimir--Polder and enantiomer-dependent decay-rate literature; we detail it, and discuss why we regard it as secondary pending a realizable low-loss design, in the End Matter.

\textit{Discussion.}---We have shown that the absence of {\it loss-free axial torque on an isotropic
particle} is not a special feature of the free-space sphere but an exact identity for any photonic
environment, including chiral and substrate-coupled ones, and that {\it shape anisotropy is the unique
escape route available at dipole order}. The resulting torque carries a strong, 
{\it sign-sensitive}
dependence on the particle's own chirality that a sphere cannot reproduce, and that a planar gold
film---used as a height-tunable near-field and standing-wave resonator rather than simply a
near-contact enhancer---can bring within reach of established torque-wrench sensitivity. The
enantiomer-differential response is structurally tied to the theorem's escape condition, and so
offers a {\it loss-free, mechanical channel for chiral-particle discrimination}, complementary to
existing lateral-force-based enantioselective
sorting~\cite{wang_lateral_2014,hayat_lateral_2015,shi_chirality-assisted_2020}.
A dedicated study of finite-thickness films, where symmetric and antisymmetric (long- and
short-range) surface-plasmon modes become accessible, may offer a further route to enhancement
beyond the broadband mechanism identified here for a bulk (half-space) substrate. Polarization-converting (wave-plate-like) metasurfaces~\cite{pang2009}, whose anisotropic reflection reshapes both the local spin density and the dressing tensor, offer a complementary route to controlling the torque, though by the identity above they remain subject to the same isotropic-particle constraint.

\begin{acknowledgments}
The authors acknowledge support from the Natural Sciences and Engineering Research Council of Canada (NSERC) Discovery Grants (R.G. and S.H.), the Canadian Foundation for Innovation (CFI), and Queen's University, Canada. 
% The authors also acknowledge the 
%  use of 
%  Claude AI for some assistance with 
%  technical derivations and manuscript preparation. 

\end{acknowledgments}

\bibliography{refs}

\clearpage
\newpage
 
\appendix
\section*{End Matter}

\textit{Explicit polarizability matrices.}---For the {\it isotropic} (spherical) particle of the no-go
theorem, the bare $6\times6$ bi-isotropic polarizability entering Eq.~\eqref{eq:tau} is
\begin{equation}
{\bm \alpha}=\begin{bmatrix}
{\boldsymbol{\alpha}}_{\rm e} & i{\boldsymbol{\alpha}}_{\rm \chi}\\
-i{\boldsymbol{\alpha}}_{\rm \chi} &{\boldsymbol{\alpha}}_{\rm m}
\end{bmatrix}
=\begin{bmatrix}
{{\alpha}}_{\rm e}\mathcal{I}_3 & i{{\alpha}}_{\rm \chi}\mathcal{I}_3\\
-i{{\alpha}}_{\rm \chi}\mathcal{I}_3 &{{\alpha}}_{\rm m} \mathcal{I}_3
\end{bmatrix},
\label{eq:alpha6}
\end{equation}
with $\mathcal{I}_3$ the $3\times3$ identity and, writing
$\mathcal{D}\equiv(\epsilon_p+2\epsilon_B)(\mu_p+2\mu_B)-\kappa^2$,
\begin{align}
\alpha_{e}&=3V_p\,\frac{(\epsilon_p-\epsilon_B)(\mu_p+2\mu_B)-\kappa^2}{\mathcal{D}}, \nonumber\\
\alpha_{m}&=3V_p\,\frac{(\mu_p-\mu_B)(\epsilon_p+2\epsilon_B)-\kappa^2}{\mathcal{D}}, \nonumber\\
\alpha_{\chi}&=9V_p\,\frac{\kappa\sqrt{\epsilon_B\mu_B}}{\mathcal{D}}.
\label{eq:alphasphere}
\end{align}
These are the $L_i\to1/3$, shape-degenerate limit of the ellipsoid polarizabilities quoted in the
main text, and they are real whenever $\epsilon_p$, $\mu_p$, and $\kappa$ are real---the condition
under which Eq.~\eqref{eq:tau} vanishes identically.

For the ellipsoid, in the ordered basis $(p_x,p_y,m_x,m_y)$, and with both transverse principal
axes in the substrate plane, the corresponding bare $4\times4$ matrix is
\begin{equation}
\boldsymbol{\alpha}=\begin{bmatrix}
\alpha_{e,1} & 0 & i\alpha_{\chi,1} & 0 \\
0 & \alpha_{e,2} & 0 & i\alpha_{\chi,2} \\
-i\alpha_{\chi,1} & 0 & \alpha_{m,1} & 0 \\
0 & -i\alpha_{\chi,2} & 0 & \alpha_{m,2}
\end{bmatrix}.
\label{eq:alpha4}
\end{equation}
Each $2\times2$ block is diagonal in the principal axes; the two magnetoelectric blocks are equal
and opposite, $\pm i\,\mathrm{diag}(\alpha_{\chi,1},\alpha_{\chi,2})$, rather than individually
antisymmetric. The shape anisotropy that escapes the no-go theorem enters only through the
inequality of the two diagonal entries within each block, $\alpha_{e,1}\neq\alpha_{e,2}$ and so
on; setting $L_1=L_2$ makes every block proportional to $\mathbf{I}_2$ and restores the exact
cancellation of Eq.~\eqref{eq:tau}.

\textit{Exact Sommerfeld substrate coupling.}---The near-field forms, Eq.~\eqref{eq:Gnearfield},
are the $k_Bh\to0$ limit of the general, fully retarded reflected Green tensor for a dipole above a
planar interface~\cite{sommerfeld1909,paulus2000,novotnyhecht2025}. Writing
$k_\rho$ for the in-plane wavevector, $k_z=\sqrt{k_B^2-k_\rho^2}$, $k_{zS}=\sqrt{\epsilon_Sk_0^2-k_\rho^2}$
(branch fixed by $\mathrm{Re}\,k_z\ge0$, $\mathrm{Im}\,k_z\ge0$, $\mathrm{Im}\,k_{zS}\ge0$; $k_0=\omega/c$), and the ordinary
Fresnel reflection coefficients $r_s=(k_z-k_{zS})/(k_z+k_{zS})$,
$r_p=(\epsilon_Sk_z-\epsilon_Bk_{zS})/(\epsilon_Sk_z+\epsilon_Bk_{zS})$ with $\mu_B=\mu_S=1$ (nonmagnetic), we evaluate
\begin{align}
g_e(h)&=\frac{ik_B^2}{8\pi}\int_0^\infty\frac{k_\rho}{k_z}\Big[r_s-\frac{k_z^2}{k_B^2}r_p\Big]e^{2ik_zh}\,dk_\rho, \nonumber\\
g_m(h)&=\frac{ik_B^2}{8\pi}\int_0^\infty\frac{k_\rho}{k_z}\Big[r_p-\frac{k_z^2}{k_B^2}r_s\Big]e^{2ik_zh}\,dk_\rho, \nonumber\\
g_c(h)&=\frac{ik_B}{8\pi}\int_0^\infty k_\rho\,(r_s-r_p)\,e^{2ik_zh}\,dk_\rho.
\label{eq:sommerfeld}
\end{align}
These are the exact coupling constants, $G_{ee}(h)\equiv g_e(h)$, $G_{em}(h)\equiv g_c(h)$, $G_{mm}(h)\equiv g_m(h)$, used
throughout the main text. 
%with $g_e\to G_{ee}(h)$, $g_c\to G_{em}(h)$ 
As $k_Bh\to0$, these reproduce the electric and
magnetoelectric near-field forms of Eq.~\eqref{eq:Gnearfield} exactly. $g_m(h)$ diverges only as
$\sim h^{-1}$, subdominant to $g_e(h)\sim h^{-3}$, which is why
Eq.~\eqref{eq:Gnearfield} has no magnetic-magnetic term; $g_m(h)$ is, however, comparable to
$g_B$ itself rather than negligible ($|g_m|/g_B\approx7,3,1.2$ at $h=13,22,52$~nm for gold), and its
imaginary (dissipative) part ${\rm Im}[g_m]$ shifts the torque of a $30$~nm lossless ellipsoid at $d_{gap}=1$~nm by $5.6\%$ (its real part by $<0.01\%$).

The integrands of $g_e$ and $g_m$ have an integrable branch-point singularity at
$k_\rho=k_B$ (the interval is split there), and all three, $g_e,g_m,g_c$, are  sharply peaked near the surface-plasmon pole of $r_p$:
for gold at $\lambda_0=1\,\mu$m, $r_p(k_\rho)$ has a surface-plasmon pole at $k_\rho/k_B=1.0217$
with linewidth $\sim10^{-3}$ (set by $\mathrm{Im}\,\epsilon_{\rm Au}$), which we resolve by
adaptive Gauss--Kronrod quadrature seeded with this location, integrating $k_\rho$ from $0$ to
$k_B\max(25,\,30/(k_Bh)+10)$.
%(evanescent contributions beyond this cutoff introduce a relative
%error smaller than $10^{-6}$).
%
Setting $\mathrm{Im}\,\epsilon_{\rm Au}\to0$ at fixed $\mathrm{Re}\,\epsilon_{\rm Au}$ does
\emph{not} remove the surface-plasmon pole: its existence condition,
$\mathrm{Re}\,\epsilon_S<-\epsilon_B$, is independent of loss, and the pole merely moves onto
the real axis at $k_\rho/k_B=1.0219$ as an undamped bound mode. The test therefore isolates
substrate dissipation rather than the mode itself. So isolated, dissipation changes the torque
by only $\sim10\%$ for $h\gtrsim140$~nm, but by $38$--$56\%$ at $d_{gap}=2$~nm for
$\kappa=0.2$ ($24$--$76\%$ at $\kappa=0$), consistent with lossy near-field channels
contributing only at near contact. The quasistatic image formula,
Eq.~\eqref{eq:Gnearfield}---no retardation, hence no pole at all---\emph{overstates} the
gold-to-silicon torque contrast, giving $11$ at $h=26$~nm rather than $3.35$.

Gold's optical constants are taken from Johnson and
Christy~\cite{johnson1972}, interpolated to $n_{\rm Au}(1\,\mu\text{m})=0.228+6.47i$ ($\epsilon_{\rm Au}=-41.8+2.9i$); silicon's from
Green~\cite{green2008}, $n_{\rm Si}(1\,\mu\text{m})=3.572+5.1\times10^{-4}i$. The near-contact torque is only weakly
sensitive to the gap: closing $d_{gap}$ from $2$~nm to $0.3$~nm raises the $60$~nm value from
$0.201$ to $0.210$~pN$\cdot$nm, a $4\%$ change, so the useful enhancement lies in the
height-optimized geometry rather than in the smallest achievable gap---which, at sub-nm
separations, would in any case fall outside the validity of the point-dipole description.

%Because scattering contributes to extinction, the extinction torque alone is nonzero whenever $P_{\rm sca}>0$, even for $P_{\rm abs}=0$; Eq.~\eqref{eq:leading} is not an absorption torque.

\textit{Generalized optical theorem and the sphere-above-substrate case.}---The identity
underlying Eq.~\eqref{eq:tau} extends further than the torque statement itself. Writing
$\tilde{\boldsymbol{\alpha}}^{-1}=\boldsymbol{\alpha}^{-1}-\mathbf{G}$ and using only that the bare $\boldsymbol{\alpha}$ is lossless
($\boldsymbol{\alpha}=\boldsymbol{\alpha}^\dagger$, real $\alpha_e,\alpha_m,\alpha_\chi$), so that $\boldsymbol{\alpha}^{-1}$ is Hermitian:
\begin{equation}
\tilde{\boldsymbol{\alpha}}^{-1}-(\tilde{\boldsymbol{\alpha}}^{-1})^\dagger=(\boldsymbol{\alpha}^{-1}-\mathbf{G})-(\boldsymbol{\alpha}^{-1}-\mathbf{G}^\dagger)=-(\mathbf{G}-\mathbf{G}^\dagger).
\end{equation}
Multiplying on the left by $\tilde{\boldsymbol{\alpha}}$ and on the right by $\tilde{\boldsymbol{\alpha}}^\dagger$, and using
$\tilde{\boldsymbol{\alpha}}\tilde{\boldsymbol{\alpha}}^{-1}=(\tilde{\boldsymbol{\alpha}}^\dagger)^{-1}\tilde{\boldsymbol{\alpha}}^\dagger=\mathbf{I}$, gives
\begin{equation}
\tfrac{1}{2i}(\tilde{\boldsymbol{\alpha}}-\tilde{\boldsymbol{\alpha}}^\dagger)=\tilde{\boldsymbol{\alpha}}\,\tfrac{1}{2i}(\mathbf{G}-\mathbf{G}^\dagger)\,\tilde{\boldsymbol{\alpha}}^\dagger,
\label{eq:optical_theorem}
\end{equation}
for \emph{any} $\mathbf{G}$---reciprocity of the environment, 
%$\mathbf{G}_{ee}{=}\mathbf{G}_{ee}^T$, $\mathbf{G}_{mm}{=}\mathbf{G}_{mm}^T$,$\mathbf{G}_{em}{=}-\mathbf{G}_{me}^T$, 
is a separate property, needed for neither Eq.~\eqref{eq:optical_theorem} nor
the torque identity Eq.~\eqref{eq:tau}. 

Equation~\eqref{eq:optical_theorem} is the dipole-order
optical theorem: the anti-Hermitian part of the dressed response is inherited entirely from the
environment, so the extinction of a lossless
particle goes entirely into radiation and substrate absorption. That the lossless particle itself
absorbs nothing follows in one line: the total field at the particle is
$\mathbf{f}=\mathbf{d}+\mathbf{G}\mathbf{u}=\boldsymbol{\alpha}^{-1}\mathbf{u}$, 
so $P_{\rm abs}\propto\mathrm{Im}\,(\mathbf{f}^\dagger \mathbf{u})=\mathrm{Im}\,(\mathbf{u}^\dagger\alpha^{-1\dagger}\mathbf{u})=0$ for Hermitian $\boldsymbol{\alpha}^{-1}$, in any environment.
This does not, by itself, forbid torque: %torque is
$\tau_z=\tfrac14\mathbf{u}^\dagger(\mathbf{T}\boldsymbol{\alpha}^{-1}-\boldsymbol{\alpha}^{-1\dagger}\mathbf{T})\mathbf{u}$ [$\mathbf{T}\equiv-\,\mathbf{I}_2\otimes \mathbf{J}$ the in-plane rotation generator of Eq.~\eqref{eq:Gmatrix} acting on
both dipole pairs], for a passive particle, vanishes identically only when
$\boldsymbol{\alpha}^{-1}$ is both Hermitian and commutes with $\mathbf{T}$; writing $\boldsymbol{\alpha}^{-1}=\mathbf{R}+i\mathbf{K}$ with $\mathbf{R},\mathbf{K}$ Hermitian, $\mathbf{T}\boldsymbol{\alpha}^{-1}-\boldsymbol{\alpha}^{-1\dagger}\mathbf{T}=[\mathbf{T},\mathbf{R}]+i\{\mathbf{T},\mathbf{K}\}$:
losslessness ($\mathbf{K}=0$) removes the absorptive channel, in-plane rotational invariance ($[\mathbf{T},\mathbf{R}]=0$) removes the reactive one, and a lossless anisotropic particle keeps the latter---the loss-free torque of this  work.

For a lossless \emph{sphere} above a substrate under normal incidence, the same conclusion
follows from angular-momentum conservation alone, independent of the polarizability algebra
above. The system is invariant under rotation about the surface normal through the sphere's
center, so every field---incident and scattered---is a $J_z$ eigenstate with the same eigenvalue
$m=\sigma$ as the incident wave (reflection reverses the helicity but not $m$); the axial
angular-momentum flux through any surface enclosing the sphere is therefore $\sigma/\omega$ times
the energy flux, and a lossless sphere, absorbing no energy, absorbs no angular momentum---for
any substrate preserving this axial symmetry, including a lossy or bulk-chiral one. 
%The axial
%symmetry supplies here exactly what $[T,H]=0$ supplies in %the algebraic argument, and exactly
%what an ellipsoid breaks.

\begin{table}[bt]
\centering
\begin{tabular}{c|cc|c}
\hline\hline
$h$ (nm) & $\tau_z^{\rm Au}$ (pN$\cdot$nm) & $\tau_z^{\rm Si}$ (pN$\cdot$nm) & ratio Au/Si \\
\hline
26  & 0.201 & 0.060 & 3.35 \\
60  & 0.196 & 0.066 & 2.97 \\
100 & 0.325 & 0.094 & 3.47 \\
150 & 0.539 & 0.151 & 3.57 \\
185 & 0.597 & 0.180 & 3.32 \\
220 & 0.532 & 0.177 & 3.01 \\
\hline\hline
\end{tabular}
\caption{Exact Sommerfeld solve, gold versus silicon substrate, same particle and illumination as
Fig.~\ref{fig:results}. The Au/Si ratio is stable to within $\pm10\%$ across an $8.5\times$ range of
height, 
confirming that little of gold's advantage can be attributed to the narrowband surface-plasmon resonance.}
\vspace{0.3cm}
\label{tab:AuSi}
\end{table}
 
\textit{Gold versus silicon, full height dependence.}---Table~\ref{tab:AuSi} extends the two-height
comparison in the main text from the near field through the first standing-wave maximum, for the same $60$/$24$~nm,
$\kappa=0.2$ ellipsoid.
 
\textit{Chiral substrate.}---For a bulk-chiral (Pasteur) half-space with scalar chirality
$\kappa_S$, permittivity $\epsilon_S$, and permeability $\mu_S=1$, the near-field ($k_\rho\to\infty$)
reflection coefficients are~\cite{butcher2012,bassiri1988}
\begin{align}
R^{(sp)}&=-R^{(ps)}=\frac{2i\kappa_S\sqrt{\epsilon_B}}{2\epsilon_B+2\epsilon_S-\kappa_S^2}, \nonumber\\
R^{(pp)}&=\frac{2\epsilon_S-\kappa_S^2-2\epsilon_B}{2\epsilon_B+2\epsilon_S-\kappa_S^2}, 
%\qquad
R^{(ss)}=\frac{-\kappa_S^2}{2\epsilon_B+2\epsilon_S-\kappa_S^2},
\label{eq:Rchiral}
\end{align}
generalizing from the vacuum-background formula of Butcher, Buhmann, and
Scheel~\cite{butcher2012} via the scaling $\epsilon_S\to\epsilon_S/\epsilon_B$,
$\kappa_S\to\kappa_S/\sqrt{\epsilon_B}$; the eigen-refractive-indices of the chiral medium
relative to the background, $n_\pm/n_B=\sqrt{\epsilon_S/\epsilon_B}\pm\kappa_S/\sqrt{\epsilon_B}$,
fix this scaling of $\kappa_S$ with $n_B$ rather than $\epsilon_B$. The antisymmetric pair
$R^{(sp)}=-R^{(ps)}$ cancels upon azimuthal integration in the ordinary electric-electric and
magnetic-magnetic self-terms, as reciprocity requires, but survives in the magnetoelectric block: the off-diagonal blocks $b\mathbf{J}$ of Eq.~\eqref{eq:Gmatrix} become
$[g_c(h)\mathbf{J}+g_\chi(h)\mathbf{I}_2]$ (upper right) and $[g_c(h)\mathbf{J}-g_\chi(h)\mathbf{I}_2]$ (lower left), the relative
sign fixed by reciprocity,
with an isotropic term $g_\chi(h)$, odd in $\kappa_S$ and scaling as $h^{-3}$ in the near
field [$g_\chi\to R^{(sp)}/(32\pi h^3)$]. 

For a representative $n_S=3.45$, 
$\kappa_S=0.2$ substrate at $h=5$~nm, $g_\chi$ is
comparable in magnitude to the achiral magnetoelectric coupling $g_c(h)$ itself ($|g_\chi|/|g_c|\approx0.6$).
For a lossless particle, however, only the anti-Hermitian part of $G$ produces torque, and for a lossless substrate the near-field $g_\chi$ is purely imaginary: the torque-relevant part is the small real part carried by the propagating spectrum, $\mathrm{Re}\,[g_\chi]\approx-0.3\,g_B$, which
modifies the
achiral-substrate torque baseline of a 10/4\,nm ellipsoid at
$d_{gap}=1$\,nm by several percent, with a sign set by $\kappa\kappa_S$. In
contrast, the $\kappa_S^2$ correction to the effective $pp$-channel reflectivity, $R^{(pp)}$ in
Eq.~\eqref{eq:Rchiral}, is a comparatively minor effect (a few times $10^{-4}$ relative, at
$\kappa_S=0.2$): the genuine substrate-chirality signature in the torque is carried almost
entirely by $g_\chi(h)$, not by the $\kappa_S^2$ renormalization of the achiral channels. 

A closed-form expression for $g_\chi(h)$ 
can be derived but is not needed explicitly; 
current values are obtained by direct numerical angular-spectrum integration of the exact $k_\rho$-dependent chiral-half-space
reflection matrix~\cite{bassiri1988}, of which Eq.~\eqref{eq:Rchiral} is the $k_\rho\to\infty$ limit. This is a genuine, independent modulation channel,
connected to the discriminatory chiral Casimir--Polder potential and enantiomer-dependent decay
rates reported previously~\cite{butcher2012,rapp2025}, but one that requires an engineered
$\kappa_S\sim0.1$--$0.2$ not found in natural bulk media; we regard it as a secondary result
pending a concrete low-loss chiral-metamaterial substrate design, rather than a claim at the same
level of experimental readiness as the achiral-substrate (gold/silicon) results above.

\end{document}